\documentclass[twocolumn,aps,prl,epsfigure,superscriptaddress]{revtex4}
\usepackage{amssymb}
\usepackage{amsmath}
\usepackage{graphicx}

\newcommand{\bra}[1]{\left\langle #1\right|}
\newcommand{\ket}[1]{\left| #1\right\rangle}

\newcommand{\ketbra}[3][]{\left|#2\right\rangle_{#1}\!\left\langle#3\right|}
\newcommand{\eeqref}[1]{Eq.~(\ref{#1})}
\begin{document}
\title{Linear-optical processing cannot increase photon efficiency}
\begin{abstract}
We answer the question whether linear-optical processing of the states produced by one or multiple imperfect single-photon
sources can improve the single-photon fidelity. This processing can include arbitrary interferometers, coherent states,
feedforward, and conditioning on results of detections. We show that without introducing multiphoton components, the
single-photon fraction in any of the single-mode states resulting from such processing cannot be made to exceed the
efficiency of the best available photon source. If multiphoton components are allowed, the single-photon fidelity cannot be
increased beyond $1/2$. We propose a natural general definition of the quantum-optical state efficiency, and show
that it cannot increase under linear-optical processing.
\end{abstract}
\author{D. W. Berry}
\affiliation{Institute for Quantum Computing, University of Waterloo, Ontario N2L 3G1, Canada}
\author{A. I. Lvovsky}
\affiliation{Institute for Quantum Information Science, University of Calgary, Alberta T2N 1N4, Canada} \maketitle

Optical implementation of quantum information processing and communication employs the single-photon state as one of its
primary resources \cite{Buller10}. There exist a variety of methods to produce this state, both of heralded and on-demand
nature \cite{Buller10,NJPissue}. However, no single-photon source is perfect. While many single-photon
sources are able to effectively suppress multiphoton components from the output, the produced state typically has a significant
admixture of vacuum. In other words, the quantum state of light generated by a typical single-photon source can be
approximately written in the photon number basis as
\begin{equation} \label{rho}
\hat\rho = (1-p)\ketbra{0}{0} + p\ketbra{1}{1},
\end{equation}
where $p$ is the efficiency of the source. In the remainder of the paper, we call state \eqref{rho} the inefficient single
photon state (ISPS).

In this work, we are investigating possibilities to enhance the efficiency of an ISPS using linear optical (LO) processing.
This processing includes arbitrary operations by means of LO elements (mirrors, beam splitters, etc.), destructive
measurements as well as modifications of the LO circuit (feedforward) or postselection based on measurement results. The
efficiency of the output ISPS is evaluated for the remaining undetected modes after postselection. 

LO processing is attractive because linear optical elements and quantum optical detectors are widely available,
inexpensive and versatile; furthermore, they can be integrated in a waveguide circuit \cite{Politi}. Recent investigations
have shown that many tasks of quantum information technology can be accomplished by means of LO processing. A well-known
example is an efficient linear optical scheme for quantum computation \cite{KLM}. Furthermore, corrections of many types
of errors, including those caused by photon loss, can be accomplished using
such processing \cite{Dawson06}. This naturally raises the question of whether LO processing can be employed to correct
for photon loss itself, thus leading to a single photon source with greater efficiency.

We find the limitations on using LO processing to improve photon sources. We give two main results, resolving
long-standing conjectures:
\begin{enumerate}
\item If no two or higher photon number components are allowed in the output
mode (which is an essential requirement for many applications, such as quantum cryptography), the probability of the
single-photon component in the output cannot be higher than the efficiency of the best single-photon source at the input,
$p_{\rm max}$.
\item If we do allow multiphoton components in the output mode, the single photon weight therein cannot exceed
the greater of $p_{\rm max}$ and $1/2$.
\end{enumerate} Both these statements are general in that they do not impose any restriction on the number of photon sources
available, the configuration of the LO scheme or destructive measurements involved.

These results were hypothesized in Refs.~\cite{Berry04b,Berry07}, which provided numerical evidence and gave proofs in some
special cases. Tightness of these results has also been demonstrated. Refs.~\cite{Berry04a,Berry04b} provided a scheme that
can increase the single-photon probability, provided multiphoton components are allowed in the output. This scheme worked if
and only if the initial single-photon probability was less than $1/2$. In a related work \cite{Berry06}, it was shown that
if there is initial coherence between the vacuum and single-photon components, then the single-photon probability can be
increased, but at the expense of the coherence. For these partially-mixed states a generalized efficiency was defined, which
cannot increase through LO processing of a single source.

One restriction on the schemes we study is that the measurements are destructive. If one could perform a quantum nondestructive
(nondemolition) measurement of the number of photons, then postselecting on detection of a single photon would improve the
efficiency. However, such measurements require nonlinearity \cite{Imoto85}, and are not performed by standard photodetectors.
It is possible to achieve an effective nondemolition measurement using linear optics and destructive measurements provided that
perfect single photons are given as a resource \cite{Kok02}. However, we cannot make use of this possibility because we require
that all available single-photon sources are imperfect with quantum efficiency no higher than $p_{\rm max}$.

Aside from the above restriction, our results are valid for arbitrary generalized quantum measurements. This includes the case
when some of the modes are not measured at all, thus accounting for optical losses or imperfect mode matching on beam splitters.

\begin{figure}[!b]
\center{\includegraphics[width=5cm]{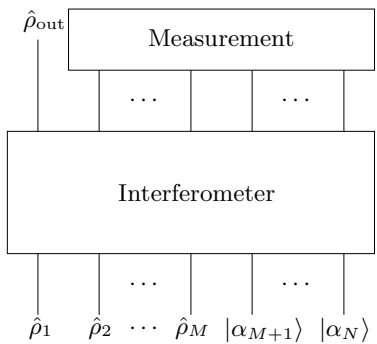}} \caption{\label{fig:int1} A general interferometer for processing
single-photon sources. The single-photon sources are the $M$ states $\hat\rho_1,\ldots,\hat\rho_M$. In addition, coherent states may
be allowed as inputs. The modes pass through a general interferometer, and all but one of the output modes are detected via a
measurement.}
\end{figure}

In general, LO processing schemes can involve feedforward, i.e.\ LO operations that are controlled by the results of measurements.
This is used, for example, in schemes for linear optical quantum computation \cite{Dawson06}. Typically, the controlled
operations are adjusted before they act on the photons.
A scheme with such feedforward can however be replaced by a scheme that is immediately prepared in its final configuration
corresponding to the set of measurement results that give the largest single-photon probability at the output. If postselection
on this set of results is employed, the single-photon probability under this fixed scheme will be at least as high as under the
scheme with feedforward. Therefore we can without loss of generality eliminate feedforward from our future analysis.

In view of the above, any LO processing scheme can be converted to the form depicted in Fig.~1 \cite{Berry04b}.
Multiple ISPSs $\hat\rho_1,\ldots,\hat\rho_M$ with efficiencies $p_1,\ldots,p_M$, are combined in a linear
interferometer. For added generality, we also allow coherent states (which are readily available from a laser) as inputs.
This includes vacuum states as coherent states with amplitude zero. All the interferometer outputs except one are then
subjected to a measurement. Conditioned on a particular result of this measurement, we analyze the state in the remaining mode. 

\begin{figure}[!b]
\center{\includegraphics[width=5cm]{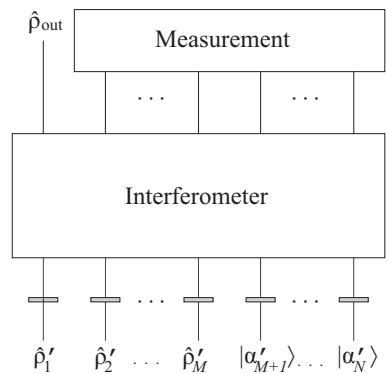}} \caption{\label{fig:int2} An equivalent scheme to Fig.~1. All interferometer
inputs of Fig.~1 are interpreted as optical states with a higher single-photon probabilities (for inefficient photon sources)
or higher amplitudes (for coherent states) that have propagated through identical attenuators (grey boxes) of transmissivity
$p_{\rm max}$.}
\end{figure}

\begin{figure}[!b]
\center{\includegraphics[width=5cm]{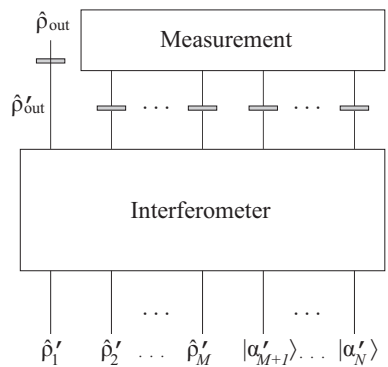}} \caption{\label{fig:int3} An equivalent scheme to Fig.~2. The interferometer
and attenuators can be interchanged, as shown in the text.}
\end{figure}

We begin our argument by redrawing our scheme as shown in Fig.~2. Because the efficiencies of all initial ISPSs are not greater
than $p_{\rm max}$, they can be interpreted as ISPSs $\hat\rho'_i$ of efficiency $p_i/p_{\rm max}$ that have propagated through
optical attenuators of transmissivity $p_{\rm max}$. The coherent states $\ket{\alpha_i}$ can be interpreted as coherent
states of amplitudes $\alpha'_i=\alpha_i/\sqrt{p_{\rm max}}$ that have been similarly attenuated.

Now the following observation can be made. \emph{The set of attenuators can be commuted with the interferometer without
affecting the multimode optical state before the measurement}. In other words, the scheme in Fig.~2 is equivalent to that in
Fig.~3. To our knowledge, this observation has been made for the first time by Varnava {\it et al.}~\cite{Varnava} for a specific
interferometer configuration. Below, we offer a general proof for an arbitrary LO process.

A multimode state propagating through an attenuator (with equal loss in all the modes) can be described as fictitious time
evolution \cite{WallsMilburn}:
\begin{equation}\label{evo}
\frac{d\hat \rho}{dt} = \kappa[\hat A (\hat \rho)- \hat N  (\hat \rho)],
\end{equation}
where
\begin{equation}
\hat A  (\hat \rho) = \sum_k \hat a_k \hat \rho \hat a_k^\dagger,
\end{equation}
\begin{equation}
\hat N  (\hat \rho) = \sum_k (\hat a_k^\dagger \hat a_k \hat \rho + \hat \rho \hat a_k^\dagger \hat a_k)/2,
\end{equation}
are the Lindbladian superoperators. Summation is performed over all optical modes with annihilation operators $\hat a_k$,
$\kappa$ is the loss coefficient such that $p_{\rm max}=e^{-\kappa t_0}$, and $t_0$ is the fictitious time during which the loss channel
is applied. In writing \eeqref{evo} we have used the fact that the attenuation is the same in all the modes.

The action of the interferometer can be written in the Heisenberg picture as a unitary transformation between the
annihilation operators of the input $\hat a_k$ and output $\hat b_k$ modes:
\begin{equation}
\hat a_j = \sum_{k} U_{jk} \hat b_k.
\end{equation}
Substituting this into the expression for $\hat A$, we get
\begin{align}
\hat A (\hat \rho) & =  \sum_j \left(\sum_{k} U_{jk} \hat b_k \right) \hat \rho \left(\sum_{l} U_{jl} \hat b_l
\right)^\dagger \nonumber \\ &= \sum_{jkl} U_{jk} U_{jl}^* \hat b_k  \hat \rho \hat b_l^\dagger \nonumber \\ &= \sum_{kl}
\delta_{kl} \hat b_k \hat \rho \hat b_l^\dagger \nonumber \\ &= \sum_k \hat b_k \hat \rho \hat b_k^\dagger.
\end{align}
In other words, the superoperator $\hat A$ has the same expression in terms of $\hat b_k$ as it does in terms of $\hat a_k$.
The same identity is valid for $\hat N$. Therefore,
equal attenuation of all the modes before the interferometer is exactly equivalent to the same attenuation after the
interferometer.

Our scheme is therefore equivalent to that shown in Fig.~3. Because the attenuators act independently on each mode, we can
without loss of generality assume the loss on mode 1 occurs \emph{after} measurement of other modes. We call the state of
mode 1 conditioned on the desired measurement result, but before the attenuation, $\hat\rho'_{\rm out}$. The diagonal
elements of density matrices $\hat\rho'_{\rm out}$ and $\hat\rho_{\rm out}$ are related by the Bernoulli transformation
\cite{Leonhardt}:
\begin{equation}\label{Bernoulli}
\bra n\hat\rho_{\rm out}\ket n=\sum\limits_{m=n}^{\infty}p^n_{\rm max}(1-p_{\rm max})^{m-n}{m \choose n} \bra
{m}\hat\rho'_{\rm out}\ket {m}.
\end{equation}
According to the above equation, if we require that state $\hat\rho_{\rm out}$ contains no photon number components with
$n>1$, the same must be true for  $\hat\rho'_{\rm out}$. The probabilities of 1-photon components in these states are then
related as
\begin{equation}\label{pr1}
X\equiv\bra 1\hat\rho_{\rm out}\ket 1=p_{\rm max}  \bra 1\hat\rho'_{\rm out}\ket 1.
\end{equation}
That is, the probability of 1 photon after the loss cannot be larger than $p_{\rm max}$. This solves the first of the two
conjectures formulated in the beginning of this paper.

We solve the other conjecture in two steps. First, we show that it is impossible to increase the probability of a single
photon above $p_{\rm max}$ for $p_{\rm max}\ge 1/2$, allowing multiphoton components in the output. For $n=1$,
\eeqref{Bernoulli} takes the form
\begin{equation}
\label{eq:sol} X = p_{\rm max} \sum_{m=1}^{\infty} (1-p_{\rm max})^{m-1} m \bra {m}\hat\rho'_{\rm out}\ket {m}.
\end{equation}
For $p_{\rm max}\ge 1/2$, $(1-p_{\rm max})^{m-1} m \le 1$, so
\begin{equation}
X \le p_{\rm max} \sum_{m=1}^{\infty} \bra {m}\hat\rho'_{\rm out}\ket {m} \le p_{\rm max},
\end{equation}
as required.

Second, we show that one cannot construct a scheme that generates the output state with the single-photon probability $X>1/2$
from inefficient SPs with $p_{\rm max}<1/2$. Indeed, suppose such a scheme exists. But then we could also use it with
inefficient SPs of any efficiency $Y$, such that $1/2<Y<X$, by first attenuating them. This would however lead to efficiency
improvement from $Y$ to $X$, which, as we just showed, is impossible.

The generality of our method also enables us to derive new results that were not anticipated in previous work. In particular,
we can define a more general form of the efficiency that may be used for states with multiphoton components. We do this by
again considering an initial state followed by a loss channel. We can define $E(\hat\rho)$ by
\begin{equation}\label{mostgen}
E(\hat\rho) \equiv \min \{ p\ |\ \exists \hat \rho^0\ge 0 ~:~ {\cal E}_p (\hat \rho^0)=\hat \rho \},
\end{equation}
where ${\cal E}_p$ indicates the loss channel with transmission probability $p$. That is, the efficiency is the minimum
transmission probability for a loss channel such that $\hat\rho$ can be obtained from a valid quantum state (i.e., with positive
semidefinite density operator).

This definition may be used for both single-mode and multimode states. In the case where the state is a tensor product of
states in the individual modes, $\hat\rho=\otimes_k \hat\rho_k$, the generalized efficiency is the maximum of that for the
individual modes: $E(\hat\rho)=p_{\rm max}\equiv\max_k(p_k)$, where $p_k=E(\hat\rho_k)$. To show this, let us define for each
mode the state $\hat\rho^0_k$, such that ${\cal E}_{p_k} (\hat \rho^0_k)=\hat \rho_k$. Then for the states
$\hat\rho'_k={\cal E}_{p_k/p_{\rm max}}(\hat\rho_k^0)$, we have
\begin{equation}
{\cal E}_{p_{\rm max}} \left(\otimes_k \hat \rho'_k\right) = \otimes_k \hat \rho_k = \hat\rho,
\end{equation}
which means, according to definition \eqref{mostgen}, that  $E(\hat\rho)\le p_{\rm max}$.
On the other hand, for any state $\hat\rho_0$ satisfying ${\cal E}_{E(\hat\rho)} (\hat \rho^0)=\hat \rho$, tracing over all
modes except $k$ gives a state
$\hat\rho''_k$ such that ${\cal E}_{E(\hat\rho)} (\hat \rho''_k)=\hat \rho_k$. Comparing this with \eeqref{mostgen}, we obtain
$\forall k\ E(\hat\rho)\ge p_k$, and thus $E(\hat\rho)=p_{\rm max}$.

By applying the procedure of commuting the loss channel with the interferometer, we find that \emph{the generalized efficiency
cannot be increased under LO processing}. This general result includes the above no-go results for improving the ISPS
efficiency as particular cases. This is because  the generalized efficiency of an ISPS is identical to that defined by
\eeqref{rho}, and for a set of ISPSs, the generalized efficiency is the maximum of that for the individual single photon sources.

A subtlety is that, with multiphoton components, the generalized efficiency does not necessarily equal the single-photon
probability. For coherent states, for example, the generalized efficiency is zero, even though the single-photon efficiency is
nonzero. For the two-photon Fock state the relation is opposite. In cases where the single-photon probability of an ISPS is
increased through LO processing (as in Refs.\ \cite{Berry04a,Berry04b}), the generalized efficiency is not increased because
multiphoton components are introduced.

An important feature of the generalized efficiency is that, if it is possible to interconvert between two states using LO
processing and postselection, then they must have the same generalized efficiency. For example, consider the case of partially
mixed states of zero and one photon,
\begin{equation}
\label{eq:part} \hat\rho =  (1 - p)\ketbra 00  + q\ketbra 01 + q^*\ketbra 10 + p \ketbra 11.
\end{equation}
It is shown in Ref.\ \cite{Berry06} that these states can be interconverted, using LO processing and conditional
measurements, with ISPSs of efficiency
\begin{equation}
\label{eq:srqeff} E'(\hat\rho)=p/(1-|q|^2/p).
\end{equation}
Hence the generalized efficiency $E(\hat\rho)$ of the partially mixed state \eqref{eq:part} must be equal to $E'(\hat\rho)$
\footnote{In fact, Ref.\ \cite{Berry06} proposed the quantity $E'(\hat\rho)$ as the ``generalized efficiency'' of the partially
mixed state \eqref{eq:part}. The new generalized efficiency we define here subsumes that earlier definition.}.

Thus we find that the technique of commuting an equal-loss channel with the interferometer enables us to resolve two
long-standing problems from previous work. We find that the efficiency of single-photon sources cannot be increased using
linear optics and destructive conditional measurements if it is required that the generated state contain no multiphoton
components. Even if this restriction is lifted, it is not possible to increase the single photon probability if $p_{\rm max}
\ge 1/2$. We formulate a general definition of the quantum efficiency of an optical state which cannot increase under LO
processing. These results place strong performance bounds on all linear optical quantum processing schemes.

One possibility that our results do not rule out is catalytic improvement of photon sources. That is, if there is one source
with very high efficiency, it might be possible to use this source to improve the efficiency of multiple sources. This is a
topic for future study.

\acknowledgments
This work has been supported by NSERC, AIF, CIFAR and Quantum{\it Works}. We thank B. C. Sanders for inspiring discussions.

\end{document}